\documentclass[preprintnumbers,prd,onecolumn,floatfix,superscriptaddress, nofootinbib]{revtex4-2}
\usepackage{graphicx}
\usepackage{setspace}
\usepackage{subfig}
\usepackage{epsfig}
\usepackage{bm}
\usepackage{amssymb}
\usepackage{float}
\usepackage{amsmath}
\usepackage{dcolumn}
\usepackage[colorlinks]{hyperref}
\usepackage[usenames,dvipsnames]{color}
\usepackage{enumitem}

\hypersetup{ breaklinks=true, pdfstartview={FitH}, colorlinks=true, linkcolor=blue, citecolor=red, filecolor=magenta, urlcolor=blue, anchorcolor=green, linktocpage=true }

\begin{document}

\title{ The cosmological model in $ f(R,T^\phi) $ gravity with Scalar Field conformity}
\author{J.~K. Singh}
\email{jksingh@nsut.ac.in}
\affiliation{Department of Mathematics, Netaji Subhas University of Technology, New Delhi-110 078, India}
\author{Akanksha Singh}
\email{akanksha.ma19@nsut.ac.in}
\affiliation{Department of Mathematics, Netaji Subhas University of Technology, New Delhi-110 078, India}
\author{Shaily} 
\email{shaily.ma19@nsut.ac.in}
\affiliation{Department of Mathematics, Netaji Subhas University of Technology, New Delhi-110 078, India}
\author{J. Jena}
\email{jjena67@rediffmail.com}
\affiliation{Department of Mathematics, Netaji Subhas University of Technology, New Delhi-110 078, India}

\begin{abstract}
\begin{singlespace}
	
\noindent The homogeneous and isotropic cosmological model in generalized $ f(R,T^\phi) $ theories associated with scalar field is discussed, which is motivated by the $ f(R,T) $ theory of gravity studied by Harko et al. \cite{Harko:2011kv, Harko:2014pqa}. The $ f(R,T^\phi) $ gravity can be explained as $ f(R,T) $ gravity with a self-interacting scalar field $ \phi $, where $ T^\phi $ is the trace of the energy-momentum tensor.  The parametrization of Hubble parameter $ H(t) $ is taken as $ \alpha-\beta e^{-\gamma t} $, where $ \alpha $, $\beta$ and $\gamma$ are arbitrary constants such that $ \alpha, \gamma>0 $ and $ \beta<0 $. The model shows no space-time singularity and the expansion of the universe will continue forever, i.e., the future scenario of the universe attains Big Freeze. The model predicts the moderate inflationary scenario at the time of the evolution of the universe and it is consistent with $ \Lambda $CDM in late times. The consistency of the model has also been examined using recent observational Hubble dataset and supernovae dataset. Finally, the physical features of the model have been discussed in some detail.  
   
\end{singlespace}

\end{abstract}

\maketitle


\section{ Introduction}\label{Intro section}

\qquad The cosmic triangle provided a format for depicting the universe's past, present, and future states. The creation of structures in the universe during the era when matter predominated necessitates a period of decelerated expansion in which gravity ought to be the dominant factor. Cosmic acceleration is a late-time occurrence. The acceleration suggests the presence of cosmic dark energy, which accelerates the expansion by outpacing the gravitational self-attraction of matter. The current accelerated expansion phase of the universe is indicated by the findings from cosmic microwave background radiation, large scale structure, and supernovae surveys \cite{SupernovaSearchTeam:1998fmf, SupernovaCosmologyProject:1998vns, WMAP:2003elm, WMAP:2008lyn, SNLS:2005qlf}. By using only supernova data, Riess et al. \cite{SupernovaSearchTeam:2004lze} presented proof for the presence of a shift from deceleration to acceleration with a certainty of greater than $ 99\% $.

The dynamics of the universe may be considerably influenced by exotic types of energy, even though it is known that the universe contains a substantial portion of ordinary matter, which slows the expansion. General relativity can account for cosmic acceleration by proposing the inexplicable cosmic fluid known as dark energy, which has a high negative pressure. The Einstein cosmological constant $ \Lambda $ \cite{Carroll:1991mt, SupernovaSearchTeam:1998bnz}, which captures the contribution of vacuum energy to the spatial curvature of space-time, is apparently the most obvious choice for dark energy. Some authors have used the reconstruction formula to describe the dark energy models \cite{Nojiri:2006be, Matsumoto:2010uv}. The apparent magnitude-redshift relationship has been discussed using Type Ia supernovae \cite{SupernovaSearchTeam:1998bnz, SupernovaCosmologyProject:1997zqe}. The basic underlying premise of dark energy models is that Einstein's General Relativity is the right gravity theory. However, it is possible that both cosmic acceleration and dark matter are indicators of a disruption in our knowledge of gravitational principles, prompting one to investigate the idea of modifying the Hilbert-Einstein Lagrangian. Modified gravity generally appears to be quite appealing since it provides qualitative solutions to a variety of fundamental issues about the late accelerating universe and dark energy. The study of the models related to the dark energy and dark matter concerns are the real challenges in the modern astrophysics and cosmology. The dark energy models discussed in the modified theory of gravity can provide viable solutions which are consistent with the recent observational dataset. 

The standard Einstein Lagrangian can be modified by substituting any arbitrary function of $ R $ for the scalar curvature $ R $, and this modified gravity is known as $ f(R) $ gravity. Furthermore, replacing the Ricci scalar $ R $ with the scalar torsion $ T $ yields $ f(T) $ gravity \cite{Bengochea:2008gz, Myrzakulov:2012axz}, and replacing with the gravitational constant $ G $ yields $ f(G) $ gravity \cite{Nojiri:2005jg, Nojiri:2005am, Cognola:2006eg, Brevik:2006nh, Elizalde:2006ub, DeFelice:2008wz}. Several additional modified theories of gravity, such as $ f(R, G) $ \cite{Bamba:2010wfw} and $ f(T, B) $ \cite{Bahamonde:2016cul} have recently been investigated. A wide range of phenomena can be produced from modified theories of gravity by adopting different functions. Although $ f(R) $ gravity is the most feasible alternative theory of gravity \cite{Sotiriou:2008rp, Flanagan:2003iw, Allemandi:2004wn, Allemandi:2004ca, Nojiri:2006ri, Capozziello:2002rd, Carroll:2003wy, Capozziello:2006uv, Boehmer:2007kx, Boehmer:2007glt, Buchdahl:1970ynr}, it does not hold up to some observational tests, such as the solar system regime \cite{Chiba:2003ir, Erickcek:2006vf}. A much general extension of $ f(R) $ gravity could be $ f(R,S_m) $, where the matter Lagrangian $ S_m $ is a function of the energy momentum tensor's trace $ T $ and is taken as $ f(R,T) $ gravity. To account for exotic imperfect fluids and quantum effects, the term $ T $ is introduced. The $ f(R,T) $ gravity can also explain the late-time cosmic acceleration. In order to address the problems raised by $ f(R) $ gravity, some observational tests \cite{Myrzakulov:2012qp, Moraes:2016gpe} have been applied to $ f(R,T) $ gravity. One can view some notable works \cite{Harko:2011kv, Harko:2014pqa, Bhardwaj:2020gat, Yadav:2020fxc, Sharma:2019oio, Singla:2019vht, Sharma:2018ikm, Yadav:2017qso, Singh:2022eun, Alvarenga:2013syu, Houndjo:2011tu, Jamil:2011ptc, Yousaf:2016lls, Moraes:2016jyi, Alves:2016iks, Sharma:2014zya, Yousaf:2017hsq, Das:2016mxq, Singh:2014kca, Nagpal:2018uza, Myrzakulov:2012ug, Myrzakulov:2012sp, Sharif:2012gz, Chattopadhyay:2012kc} to fully comprehend the $ f(R,T) $ theory. 

Recent observations suggest the existence of a scalar field which is presently evolving on cosmological time scales. Numerous fundamental theories also report the existence of the scalar field $ \phi $, which encourages us to explore the scalar field's dynamical characteristics in cosmology. There have been many different scalar-field cosmological models proposed thus far \cite{Vilenkin:1982de, Salopek:1988qh, Khalatnikov:1992sj, Caldwell:1997ii, Khoury:2003aq, Singh:2016eom, Sami:2002se, Singh:2018tlm, Singh:2012zzf}. A dark energy component with negative pressure is causing the universe to expand at an accelerated rate right now. The dark energy component of the universe is often interpreted either in terms of a cosmological constant or as a scalar field. Even though some research suggests that dark energy is a cosmological constant, it is more generally modeled as a quintessence: a scalar field rolling down a flat potential \cite{Caldwell:1997ii, Ratra:1987rm, Copeland:2006wr, Martin:2008qp, Zimdahl:2001ar, Amendola:1999er}. Distinctively, quintessence models exhibiting `tracker' properties enable the scalar field to dominate the current universe regardless of the initial conditions, avoiding the coincidence and fine-tuning issues \cite{Zlatev:1998tr, Steinhardt:1999nw}. Johari \cite{Johri:2000yx} has introduced the concept of integrated tracking, which essentially shows that the tracker potentials follow a definite path of evolution of the universe, in compatibility with the observational constraints. In the literary works, various models of the original quintessence theory have been discussed. A few of these models include the possibility of a scalar field evolution driven by a non-canonical kinetic term \cite{Armendariz-Picon:2000nqq} and a non-minimal coupling between quintessence and dark matter \cite{Amendola:1999er, Amendola:2000uh, Gasperini:2001pc, Khoury:2003aq}. Having an unclumped form of energy density permeating the entire cosmos is the only way to achieve a flat universe and, consequently, account for the CMB measurements, if we accept the estimates of matter density derived from observations of clusters \cite{White:1993wm}. The idea that the unknown, unclumped energy is due to a scalar field $ \phi $ that has not yet reached its ground state, which is called dynamical lambda or quintessence, has received much attention \cite{Ratra:1987rm, Caldwell:1997ii, Frieman:1995pm, Coble:1996te}. In contrast to the $ \Lambda $CDM, which is static and never changes, the quintessence model is dynamic and changes over time \cite{Sami:2002se, Caldwell:1997ii}. The spintessence \cite{Boyle:2001du}, the k-essence \cite{Armendariz-Picon:1999hyi, Chiba:1999ka}, the quintom \cite{Feng:2004ad}, and the tachyon \cite{Sen:2002in, Padmanabhan:2002cp} are some additional scalar fields dark energy models that have been presented. The $ f(R,T^\phi) $ technique simply presents the $ f(R,T) $ gravity as a function of $ R $ and the trace of the energy-momentum tensor of a self-interacting scalar field $ \phi $, while the matter field energy-momentum tensor gets entered into the field equations in a general manner. By incorporating a scalar field, the $ f(R,T^\phi) $ theory was accounted for in \cite{Moraes:2016gpe}. The $ T^\phi $ term makes an appropriate contribution to the paper, and as a result, even for regimes where $ T = 0 $, the theory can be separated from the $ f(R) $ gravity.

The paper is structured as follows: In section \ref{Field equations section}, we discuss $ f(R,T^\phi) $ theory of gravity and the solutions of the Einstein field equations with some physical features of the model. Section \ref{Third section} covers some specific geometrical interpretations of the model using error bar plots, statefinder diagnostic tools, the massless scalar potential $ V(\phi) $, the quintessence-like and phantom-like scalar field kinetic term $ \dot{\phi}^2 $ during the cosmic evolution. Finally, we conclude our results in Section \ref{Conclusions section}. 

\section{ Field equations and cosmic evolution in $ f(R,T^\phi) $ gravity}\label{Field equations section}

The following action results in the gravitational field equations in $ f\left(R,T^\phi\right) $ gravity.
\begin{equation} \label{1}
\displaystyle{I= \frac{1}{2} \int \left( f\left(R,T^\phi\right) + 2 L_\phi \right) \sqrt{-g} ~dx^4,}
\end{equation}

where  $ f\left(R,T^\phi\right) = -\left(R+mR^2+nT^\phi\right) $ and $ L_\phi = -\left(\frac{\epsilon}{2} ~\phi_{,\mu} \phi^{,\mu} - V\left(\phi\right)\right) $. \\

Here, we have used $ 8\pi G=1 $. \\

We vary action in Eq. (\ref{1}) \textit{w.r.t.} $ g_{\mu \nu} $ and get the following equation
\begin{equation}\label{2}
f_R \left(R,T^\phi\right) R_{\mu \nu}-\frac{1}{2} f\left(R,T^\phi\right) g_{\mu \nu} + \left(g_{\mu \nu}\Box-\nabla_\mu \nabla_\nu \right)f_R \left(R,T^\phi\right) = T_{\mu \nu}-f_{T^\phi} \left(R,T^\phi\right) \left(T_{\mu \nu}+\Theta_{\mu \nu}\right),
\end{equation}

where $\Box$ is the d' Alembert operator defined by $\Box= g^{\mu \nu}\nabla_{\mu} \nabla_{\nu}$ and $\nabla_\mu$ indicates the covariant derivative \textit{w.r.t.} $g_{\mu \nu}$ associated with the symmetric Levi-Civita connection.

\begin{eqnarray}\label{3}
T_{\mu \nu} = -\frac{2}{\sqrt{-g}} \frac{\delta\left(\sqrt{-g} L_\phi \right)}{\delta g^{\mu \nu}} = g_{\mu \nu}L_\phi - 2\frac{\delta L_\phi}{\delta g^{\mu \nu}}
\end{eqnarray}
and
\begin{eqnarray}\label{4}
\Theta_{\mu \nu} = g^{\alpha \beta} \frac{\delta T_{\alpha \beta}}{\delta g^{\mu \nu}} = -2T_{\mu \nu} + g_{\mu \nu}L_\phi - 2 g^{\alpha \beta} \frac{\delta^2 L_\phi}{\delta g_{\mu \nu} \delta g^{\alpha \beta}}.
\end{eqnarray}

We have
\begin{eqnarray}\label{5}
L_\phi = -\left(\frac{\epsilon}{2} ~\phi_{,\mu} \phi^{,\mu} - V(\phi)\right) = -\frac{1}{2}\epsilon\dot{\phi}^2 + V(\phi),
\end{eqnarray}
where dot represents the differentiation of the function \textit{w.r.t.} cosmic time $ t $. \\

Using Eq. (\ref{5}) in Eq. (\ref{4}) we get,
\begin{equation}\label{6}
\Theta_{\mu \nu} = -2T_{\mu \nu} + g_{\mu \nu} \left(-\frac{1}{2}\epsilon\dot{\phi}^2 + V\left(\phi\right) \right).
\end{equation}

Using Eq. (\ref{6}) in Eq. (\ref{2}) we get,
\begin{equation}\label{7}
f_R \left(R,T^\phi\right) R_{\mu \nu}-\frac{1}{2} f\left(R,T^\phi\right) g_{\mu \nu} + \left(g_{\mu \nu}\Box-\nabla_\mu \nabla_\nu \right)f_R \left(R,T^\phi\right) = T_{\mu \nu}+f_{T^\phi} \left(R,T^\phi\right) \left(T_{\mu \nu} + g_{\mu \nu} \left(\frac{1}{2}\epsilon\dot{\phi}^2 - V\left(\phi\right) \right)\right).
\end{equation}

Contracting Eq. (\ref{7}) w.r.t. $ g^{\mu \nu} $ we get,
\begin{equation}\label{8}
Rf_R \left(R,T^\phi\right)-2f\left(R,T^\phi\right)+3\Box f_R \left(R,T^\phi\right) = T^\phi+f_{T^\phi} \left(R,T^\phi\right) \left(T^\phi+4\left(\frac{1}{2}\epsilon\dot{\phi}^2 - V\left(\phi\right) \right)\right).
\end{equation}

From Eqs. (\ref{7}) and (\ref{8}) we get,
\begin{equation}\label{9}
R_{\mu \nu}-\frac{1}{2}Rg_{\mu \nu} = \frac{1}{f_R \left(R,T^\phi\right)}~\left(T_{\mu \nu}+T_{\mu \nu}^{'}\right),
\end{equation}
where
\begin{equation}	
T_{\mu \nu}^{'} = f_{T^\phi} \left(R,T^\phi\right) \left(T_{\mu \nu} + g_{\mu \nu} \left(\frac{1}{2}\epsilon\dot{\phi}^2-V\left(\phi\right) \right)\right) + \frac{1}{2} \left(f\left(R,T^\phi\right)-Rf_R \left(R,T^\phi\right)\right)g_{\mu \nu} + \left(\nabla_\mu \nabla_\nu-g_{\mu \nu}\Box \right)f_R \left(R,T^\phi \right).\nonumber
\end{equation}

We have
\begin{equation}\label{10}
f\left(R,T^\phi\right)= -\left(R+mR^2+nT^\phi\right).
\end{equation}

When using the homogeneous, isotropic, $ f(R,T^\phi) $ gravity, the spatially flat FLRW model's metric is,
\begin{equation}\label{11}
ds^{2} = dt^{2}-a^{2}\left(t\right) \left(dr^2+r^2\left(d\theta^2+\sin^2\theta d\phi^2\right)\right).
\end{equation}

For this given metric,
\begin{eqnarray}\label{12}
R_{11} = 2\dot{a}^2+a\ddot{a},	\nonumber \\
R_{22} = r^2 \left(2\dot{a}^2+a\ddot{a}\right) = r^2 R_{11},	\nonumber \\
R_{33} = r^2\sin^2\theta \left(2\dot{a}^2+a\ddot{a}\right) = r^2\sin^2\theta R_{11},  \nonumber \\
R_{44} = -3\frac{\ddot{a}}{a},	\nonumber \\
R = -6\left(\frac{\dot{a}^2}{a^2}+\frac{\ddot{a}}{a}\right).	
\end{eqnarray}

The energy$-$momentum tensor of a scalar field $ \phi $ with self-interacting scalar potential $ V\left(\phi\right) $ for a perfect fluid is
\begin{equation}\label{13}
T_{\mu \nu} = \epsilon\phi_{,\mu}\phi_{,\nu} - g_{\mu \nu}\left[\frac{\epsilon}{2} \phi_{,\sigma} \phi^{,\sigma} - V\left(\phi\right)\right].
\end{equation}

Using Eq. (\ref{13}) we get the following values,
\begin{eqnarray}\label{14}
T_{11} = a^2\left(\frac{\epsilon}{2}\dot{\phi}^2  - V\left(\phi\right)\right),	\nonumber \\
T_{22} = a^2r^2\left(\frac{\epsilon}{2}\dot{\phi}^2 - V\left(\phi\right)\right) = r^2T_{11},	\nonumber \\
T_{33} = a^2r^2\sin^2\theta\left(\frac{\epsilon}{2}\dot{\phi}^2 - V\left(\phi\right)\right) = \sin^2\theta ~T_{22},	\nonumber \\
T_{44} = \frac{\epsilon}{2}\dot{\phi}^2+V\left(\phi\right),	\nonumber \\
T^\phi = -\epsilon\dot{\phi}^2+4V\left(\phi\right).	
\end{eqnarray}

Using Eqs. (\ref{9}), (\ref{10}), (\ref{11}), (\ref{12}), (\ref{14}), we obtain the following field equations
\begin{eqnarray}\label{15}
2\dot{H}+3H^2-6m\left(26\dot{H}H^2+2H\dddot{H}+12H\ddot{H}+9\dot{H}^2\right) = \left(1-n\right)\frac{\epsilon}{2}\dot{\phi}^2+\left(2n-1\right)V\left(\phi\right),	\nonumber \\
3H^2-18m\left(6\dot{H}H^2+2H\ddot{H}-\dot{H}^2\right) = \left(n-1\right)\frac{\epsilon}{2}\dot{\phi}^2+\left(2n-1\right)V\left(\phi\right).
\end{eqnarray}

Resolving equations in (\ref{15}) provides the following value for $V$ and $\dot{\phi}^2$
\begin{eqnarray}\label{16}
V\left(\phi\right) = \frac{6H^2+2\dot{H}-12m\left(22\dot{H}H^2+9H\ddot{H}+3\dot{H}^2+\dddot{H}\right)}{2(2n-1)},	\nonumber \\
\dot{\phi}^2 = \frac{-2\dot{H}+12m\left(4\dot{H}H^2+3H\ddot{H}+6\dot{H}^2+\dddot{H}\right)}{(n-1)\epsilon}.
\end{eqnarray}
 
In the context of FLRW cosmology, the energy density $ \rho_\phi $ and pressure $ p_\phi $ are given as functions of the scalar field $ \phi $ by \cite{Barrow:1988yc, Barrow:1990vx}
\begin{eqnarray}\label{17}
\rho_\phi = \frac{1}{2}\epsilon\dot{\phi}^2+V\left(\phi\right),	\nonumber \\
p_\phi = \frac{1}{2}\epsilon\dot{\phi}^2-V\left(\phi\right).
\end{eqnarray}
Each equation's right-hand side is written for a field with an arbitrary potential $ V(\phi) $. \\

Using Eqs. (\ref{16}) in (\ref{17}) we get
\begin{eqnarray}\label{18}
\rho_\phi = \frac{-n\dot{H}+3\left(n-1\right)H^2+6m\left(\left(-14n+18 \right)\dot{H}H^2+\left(-3n+6\right)H\ddot{H}+\left(9n-3\right)\dot{H}^2+n\dddot{H}\right)}{\left(n-1\right)\left(2n-1\right)},	\nonumber	\\
p_\phi = \frac{\left(-3n+2\right)\dot{H}-3\left(n-1\right)H^2+6m\left(\left(30n-26\right)\dot{H}H^2+\left(15n-12\right)H\ddot{H}+\left(15n-9\right)\dot{H}^2+\left(3n-2\right)\dddot{H}\right)}{\left(n-1\right)\left(2n-1\right)}.
\end{eqnarray}

The parametrized form of Hubble parameter $ H $ used to get the solution of the field equations is as follows:
\begin{equation}\label{19}
H(t) = \alpha-\beta e^{-\gamma t},
\end{equation}
where we have arbitrary constants $\alpha$, $\beta$ and $\gamma$ such that $ \alpha>0 $, $ \beta<0 $, $ \gamma>0 $. The motivation for using this parametrization is to discuss the inflationary scenario at the early evolution of the universe and the ultimate fate of the universe for which the current observations show that the expansion of the universe is accelerating and attains the Big Freeze condition in late times. The constraints imposed on the parametrization for the Hubble parameter $ H $ are to ensure that the overall function always returns a positive value and is a decreasing function of time $ t $.

From Eq. (\ref{19}), the cosmic scale factor $ a(t) $ and the deceleration parameter $ q(t) $ can be calculated as
\begin{equation}\label{20}
a(t) = e^{\alpha t + \frac{\beta}{\gamma}e^{-\gamma t}},  \mbox{\hspace{2em}}  q(t) = -1-\frac{\beta \gamma e^{\gamma t}}{\left(\beta -\alpha e^{\gamma t}\right)^2}.
\end{equation}

\begin{figure}\centering
	\subfloat[]{\label{a}\includegraphics[scale=0.4]{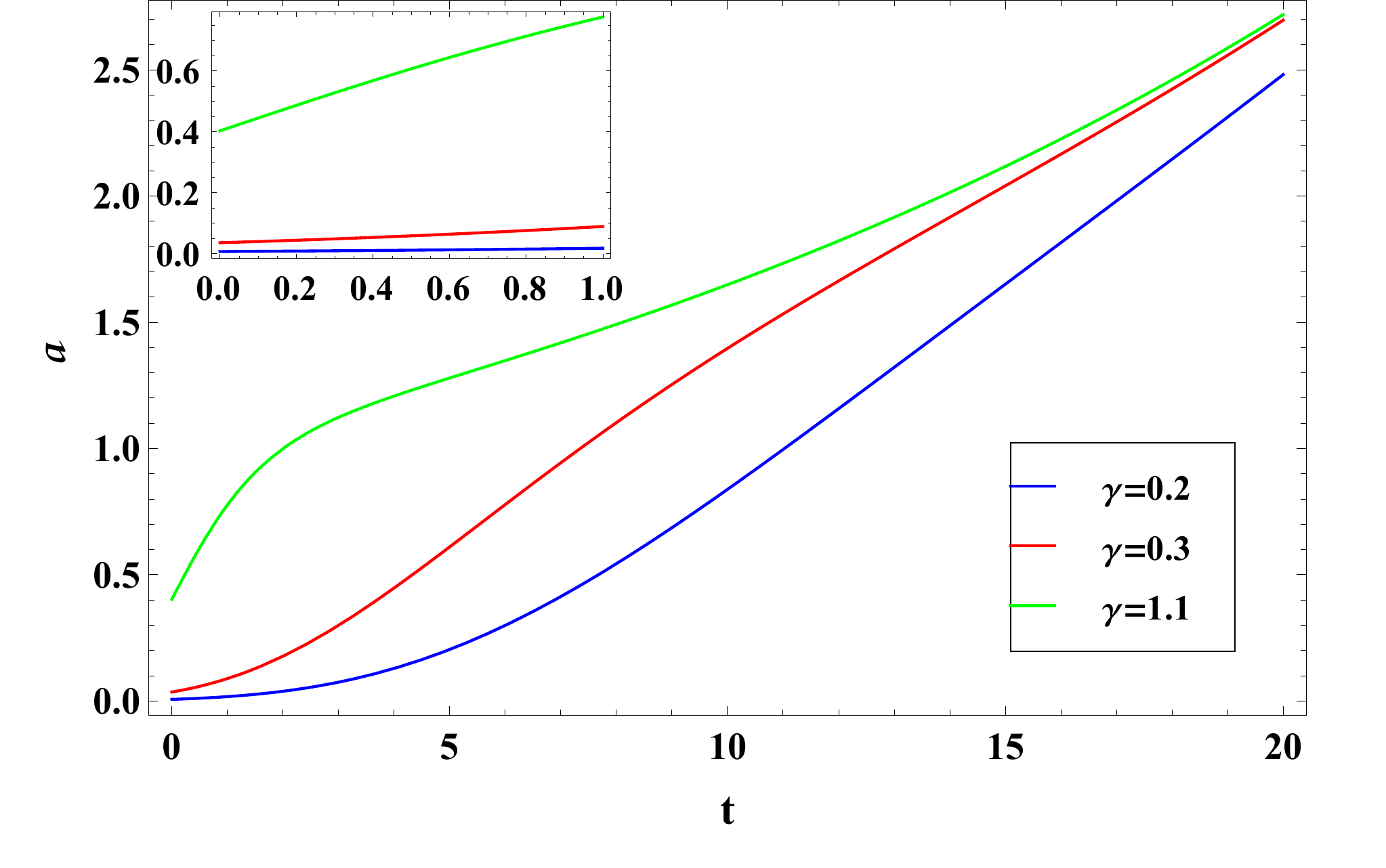}}\hfill
	\subfloat[]{\label{b}\includegraphics[scale=0.45]{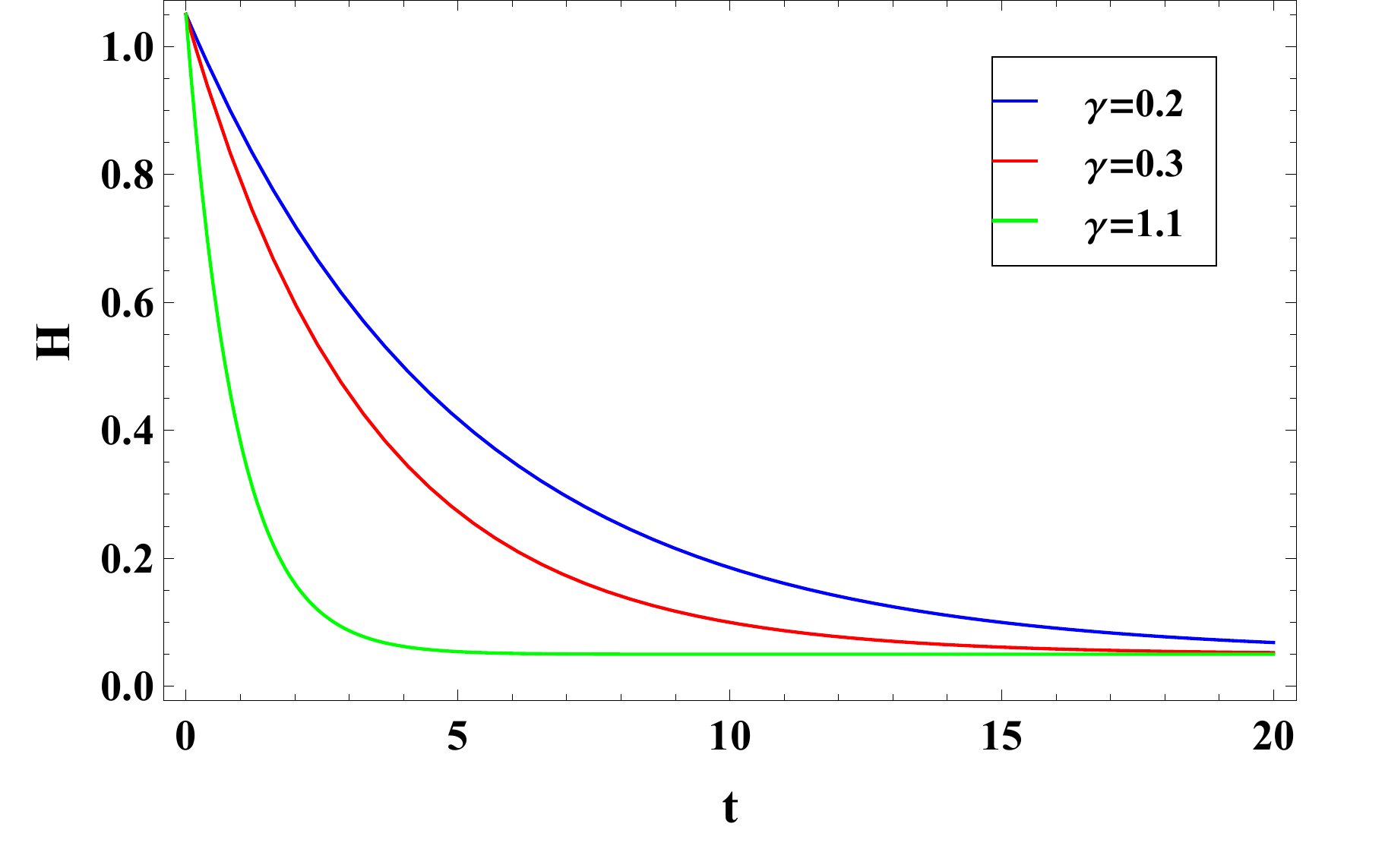}}\par 
	\subfloat[]{\label{c}\includegraphics[scale=0.45]{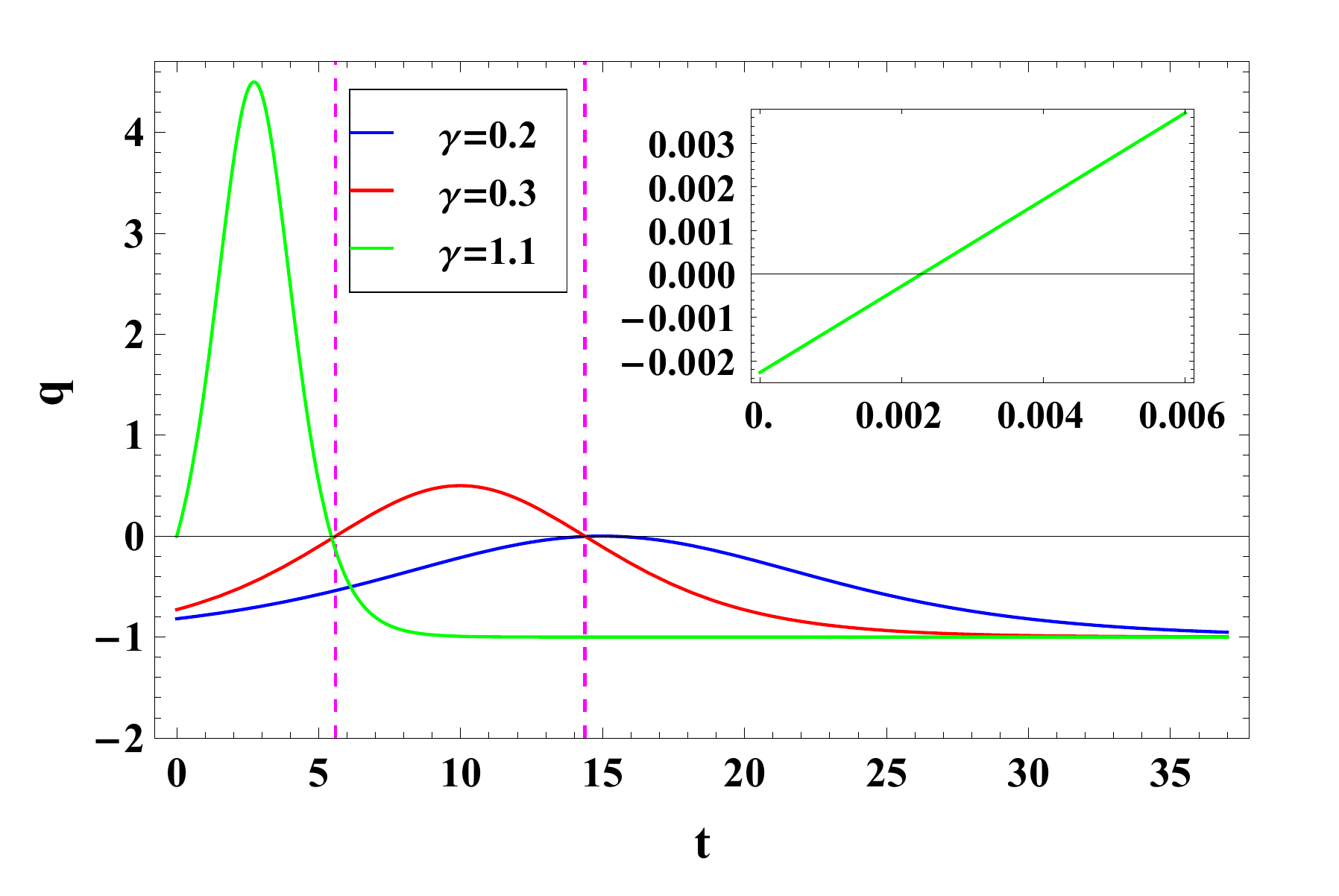}}
	\caption{\scriptsize Depiction of graphs of $ a $, $ H $ and $ q $ vs. $ t $ with $ \alpha=0.05 $ and $ \beta=-1 $. The model shows inflation at the early epoch and accelerating in late times in all cases of $ \gamma=0.2,\, 0.3,\, 1.1 $. } \label{Fig:aHq-t}
\end{figure}

Fig. \ref{Fig:aHq-t} depicts a visual representation of the scale factor $ a $, Hubble parameter $ H $, and deceleration parameter $ q $ with respect to cosmic time $ t $. We have chosen $\alpha$, $\beta$ as $0.05$ and $-1$ respectively.

According to Fig. \ref{Fig:aHq-t}(a), for $\gamma=0.2$ and $0.3$, the values of parameter $a$ increase continuously as we approach the late universe, whereas for $\gamma=1.1$, the value of scale factor $a$ increases, then decreases for some time in the early universe, and finally increases as we approach the late universe. This model is free of initial singularity because $a$ has a finite non-zero value for $t = 0$ for all of the $\gamma$ values mentioned, and this is easily observable for $\gamma=0.3 $ and $1.1 $ from the plot embedded in the Fig. \ref{Fig:aHq-t}(a), while for $\gamma=0.2 $, it can be easily verified with the help of value \[ \lim_{t\to\ 0} a(t) = 0.00673795. \]

The Hubble parameter, $ H $, aids in determining the rate of cosmic expansion. The value of $ H $ is positive for all values of $ t $ for $ \gamma=0.2, 0.3, 1.1 $ as shown in Fig. \ref{Fig:aHq-t}(b), and this suggests that the universe is expanding. As, for all values of $ \gamma $ mentioned in the model, the Hubble parameter $ H(t) $ satisfies the following property
\[ \lim_{t\to\infty} H(t) = 0.05. \]
Therefore, we can conclude that our model will experience the Big Freeze condition. 

For all values of $ \gamma $, the universe is expanding at an accelerated rate with mild inflation and attains the de-Sitter regime $ (q = -1) $ \cite{Bolotin:2015dja} in late time, as shown in Fig. \ref{Fig:aHq-t}(c). The deceleration parameter $ q $ shows transition to an accelerating state for $ \gamma = 0.3 $ and $ 1.1 $ while exhibiting eternal acceleration for $ \gamma = 0.2 $. 

The scalar field $ \phi $ is independent of the space coordinates because the metric space is homogeneous. So, we consider $ \phi $ to be a function of $ t $ only, and this can be used to get the time derivative $ d\phi / dt = \dot{\phi} $, whereas all the spatial components of the gradients, $ \partial \phi / \partial x, \partial \phi / \partial y $ and $ \partial \phi / \partial z $, are identically zero, which is consistent with the isotropic and expanding properties of the model. From symmetry analysis, it is evident that the energy density $ \rho_\phi $ and the isotropic pressure $ p_\phi $ together make up the mechanical properties of the field.

\begin{figure}\centering
	\subfloat[]{\label{a}\includegraphics[scale=0.52]{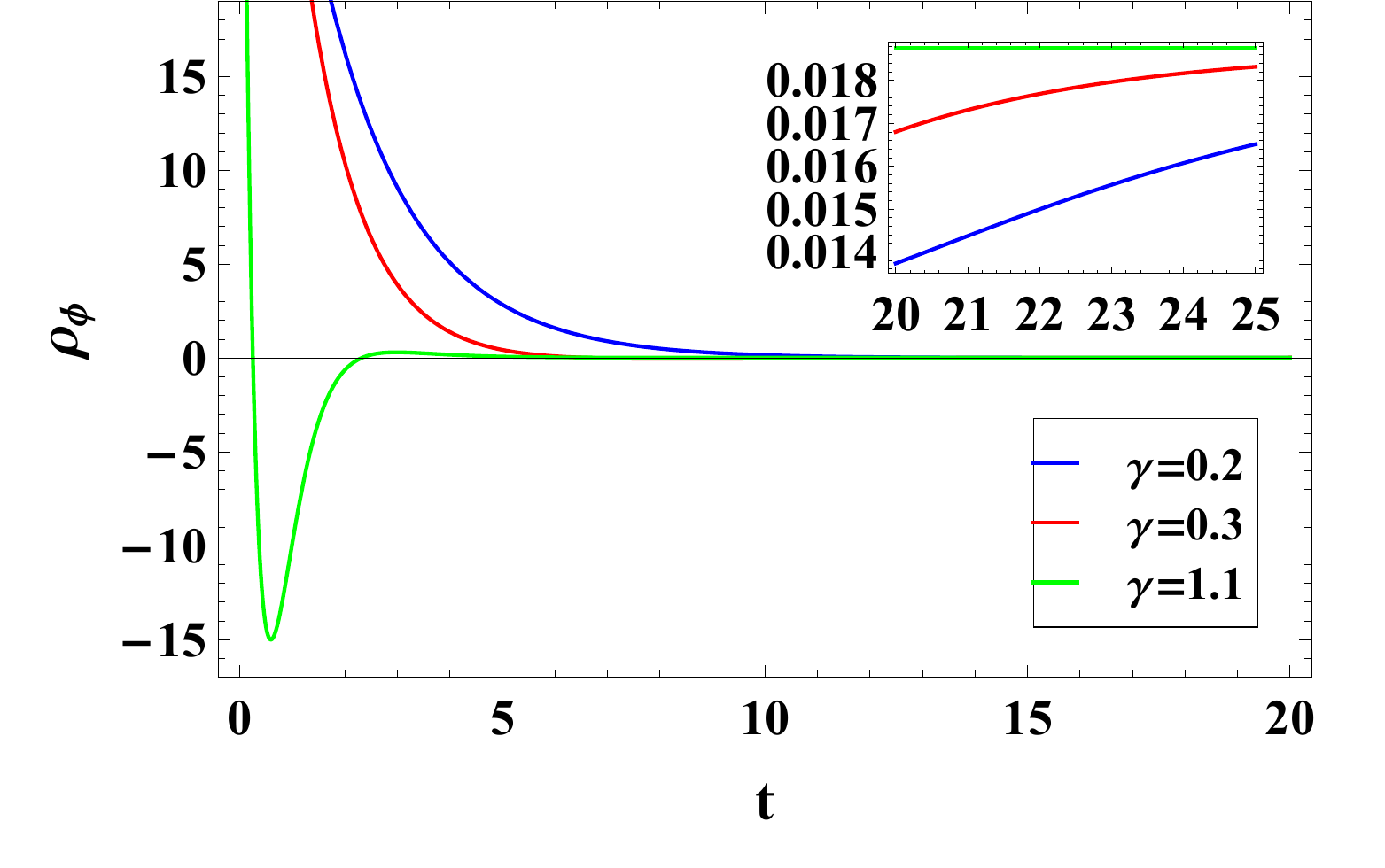}}\hfill
	\subfloat[]{\label{b}\includegraphics[scale=0.45]{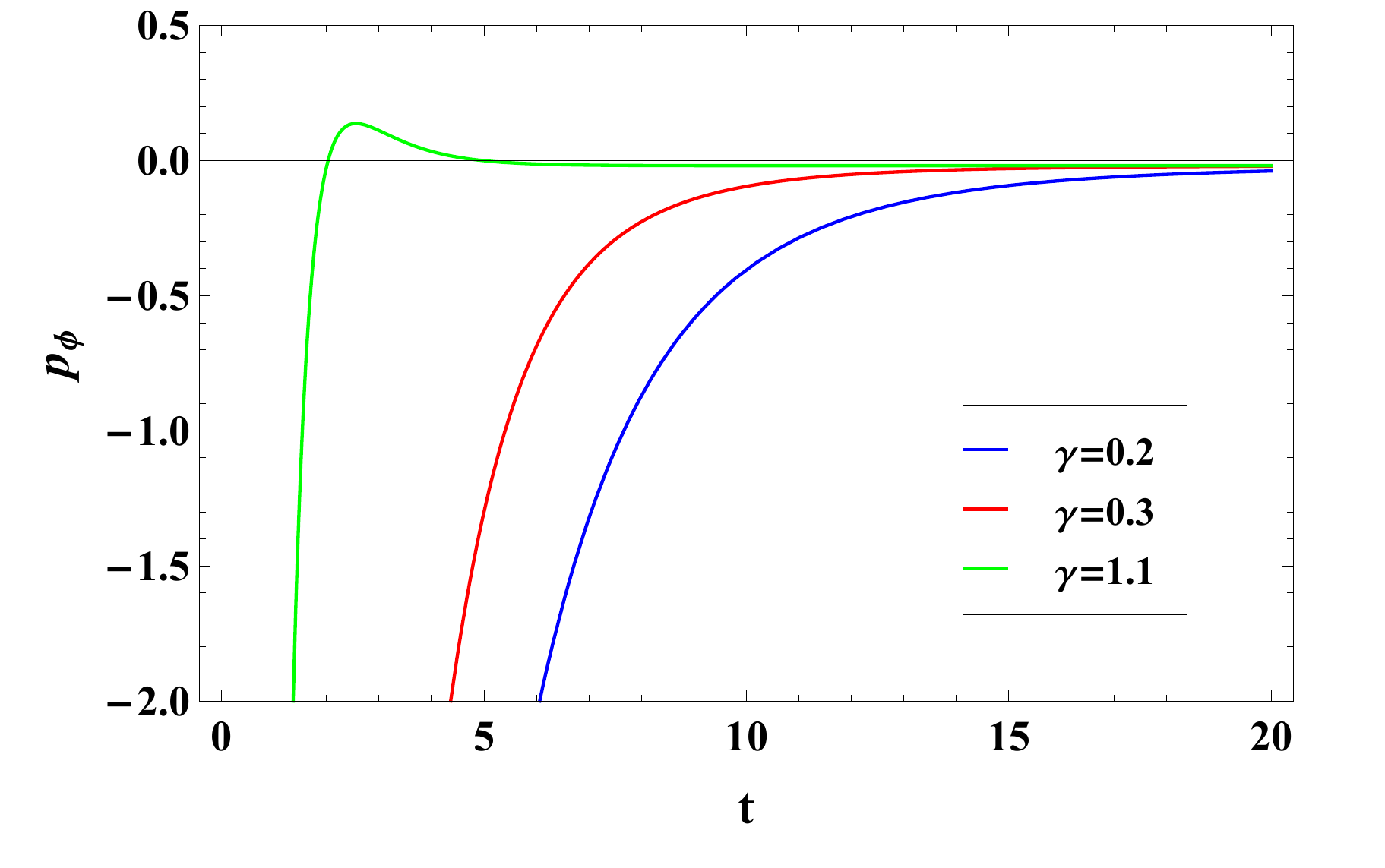}}\par 
	\subfloat[]{\label{c}\includegraphics[scale=0.48]{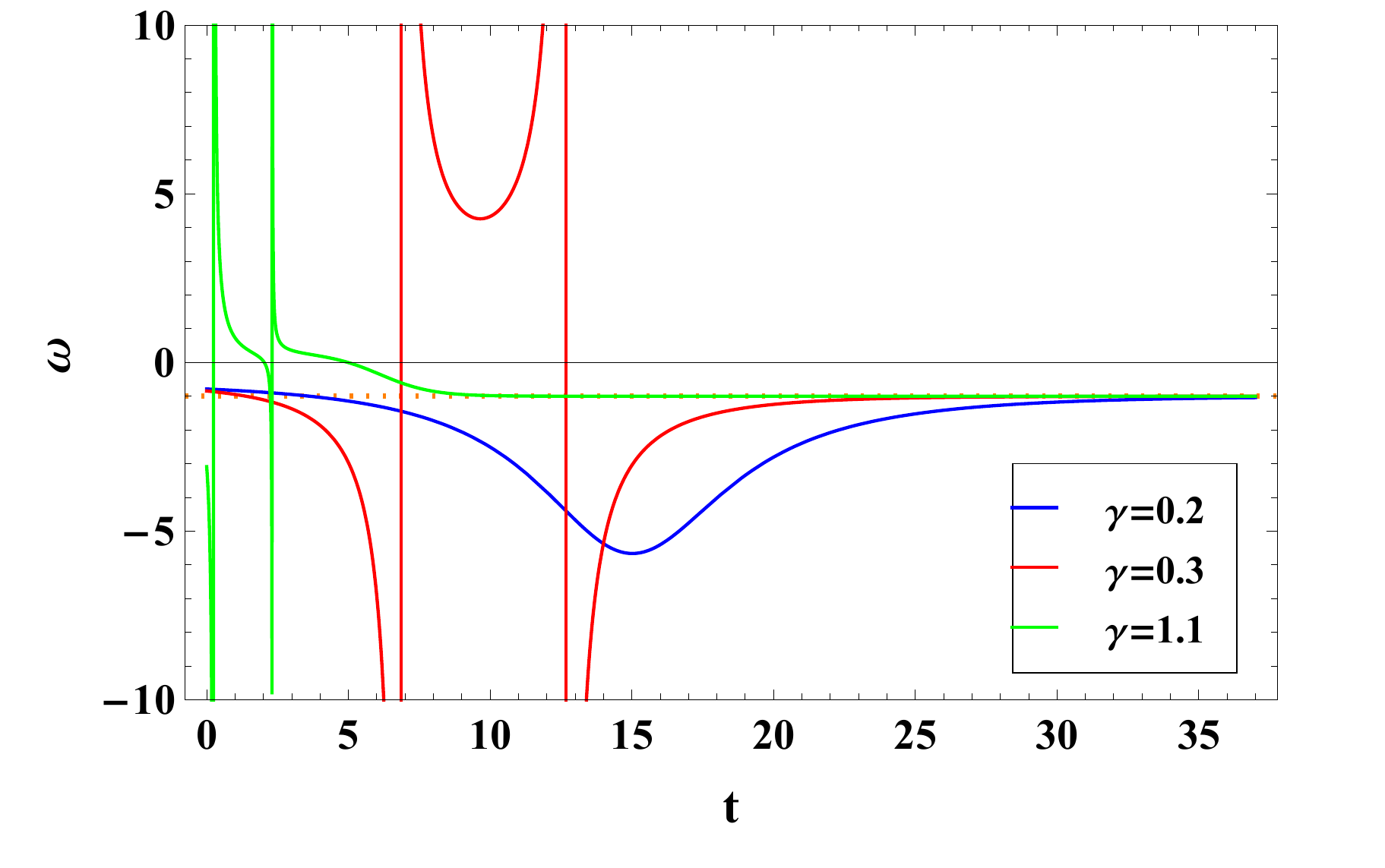}}
	\caption{\scriptsize Depiction of $ \rho_\phi-t $, $ p_\phi-t $  and $ \omega-t $ plots with $ \alpha=0.05 $, $ \beta=-1 $, $ m=0.6 $ and $ n=0.7 $. The scalar field energy density $ \rho_\phi $ and pressure $ p_\phi $ approach to finite quantity as $ t\rightarrow \infty $. The EoS parameter $ \omega \rightarrow -1 $ \textit{i.e.} the model is consistent with $ \Lambda $CDM in late times. } \label{Fig:rhopomega-t}
\end{figure}

Fig. \ref{Fig:rhopomega-t}(a) provides the variation of energy density $ \rho_\phi $ for $ \gamma=0.2, 0.3 $ and $ 1.1 $ with respect to $ t $. As time passes since the universe creation, for $ \gamma=0.2 $ and $ 0.3 $, the value of $ \rho_\phi $ falls and gets closer to a finite positive value as time $ t $ approaches $ \infty $ while, for $ \gamma=1.1 $, the value of $\rho_\phi $ decreases, reaches its minimum, and then increases for a while in the early universe, followed by a decrease in its value at present and late times of the universe, it eventually approaches the same finite positive value as for the $ \gamma=0.2, 0.3 $. Although the idea that a non-minimally coupled scalar field energy density can be negative has been understood for a long time \cite{Hiscock:1990ew, Bekenstein:1975ww, Ford:1987de, Madsen:1988ph}, the problematic ramifications have been brought to light by the works of \cite{Barcelo:2000zf, Ford:2000xg}. Here, $ \rho_\phi $ has negative values only for $ \gamma=1.1 $ for a short interval of time at the early stage of evolution. The function $ \rho_\phi $ has values of $ 52.6391 $, $ 68.4837 $, and $ 59.7011 $ for $ \gamma = 0.2 $, $ 0.3 $, and $ 1.1 $ at $ t=0 $, respectively, indicating that our model is free of initial singularity. 

The scalar field pressure $ p_\phi $ for this model is shown in Fig. \ref{Fig:rhopomega-t}(b) for all the values of $ \gamma $. The scalar field pressure $ p_\phi $ increases monotonically and approaches a finite negative value as $ t \rightarrow \infty $ in the case of $ \gamma=0.2, 0.3 $, whereas in the case of $ \gamma=1.1 $, $ p_\phi $ increases from negative to positive, reaches its maximum value, and finally decreases to the same finite negative value as for the $ \gamma=0.2, 0.3 $ in late times. Therefore, our model appears to be accelerating both now and in the near future. 

The equation of state (EoS) parameter explains the various cosmic regimes of the evolution of the universe. Using Eqs. (\ref{18}), we see the evolution of EoS parameter $ \omega =p_\phi/\rho_\phi $, for $ \gamma=0.2, 0.3 $ and $ 1.1 $ in Fig. \ref{Fig:rhopomega-t}(c). In the case of $ \gamma=0.2, 0.3 $, $ \omega $ begins in the quintessence region $ \left(-1<\omega<0\right) $, crosses the quintom line $ \left(\omega=-1\right) $, enters into the phantom region, and then approaches $ \Lambda $CDM $ \left(\omega=-1\right) $ in the late times. The model fluctuates in short intervals at the early epoch for $ \gamma=1.1 $, but later exhibits properties of a quintessence model and eventually tends towards the $ \Lambda $CDM. The model is consistent with $ \Lambda $CDM in late times for all cases. 

\section{ Dynamics of the model}\label{Third section}
\qquad We depict the error bar plots using datasets $ H(z) $ and $ SNeIa $ to examine the consistency  of our model with $ \Lambda $CDM  for the cases of $ \gamma=0.2, 0.3 $ and $ 1.1 $. Also, we use other parameters like statefinder diagnostic, scalar potential and $ \dot{\phi}^2 $, to distinguish our model from the different models and to know more about the characteristics of our model.

\subsection{ Error bar plots using $ H(z) $ and $ SNeIa $ datasets}
\begin{figure}\centering
	\subfloat[]{\label{a}\includegraphics[scale=0.47]{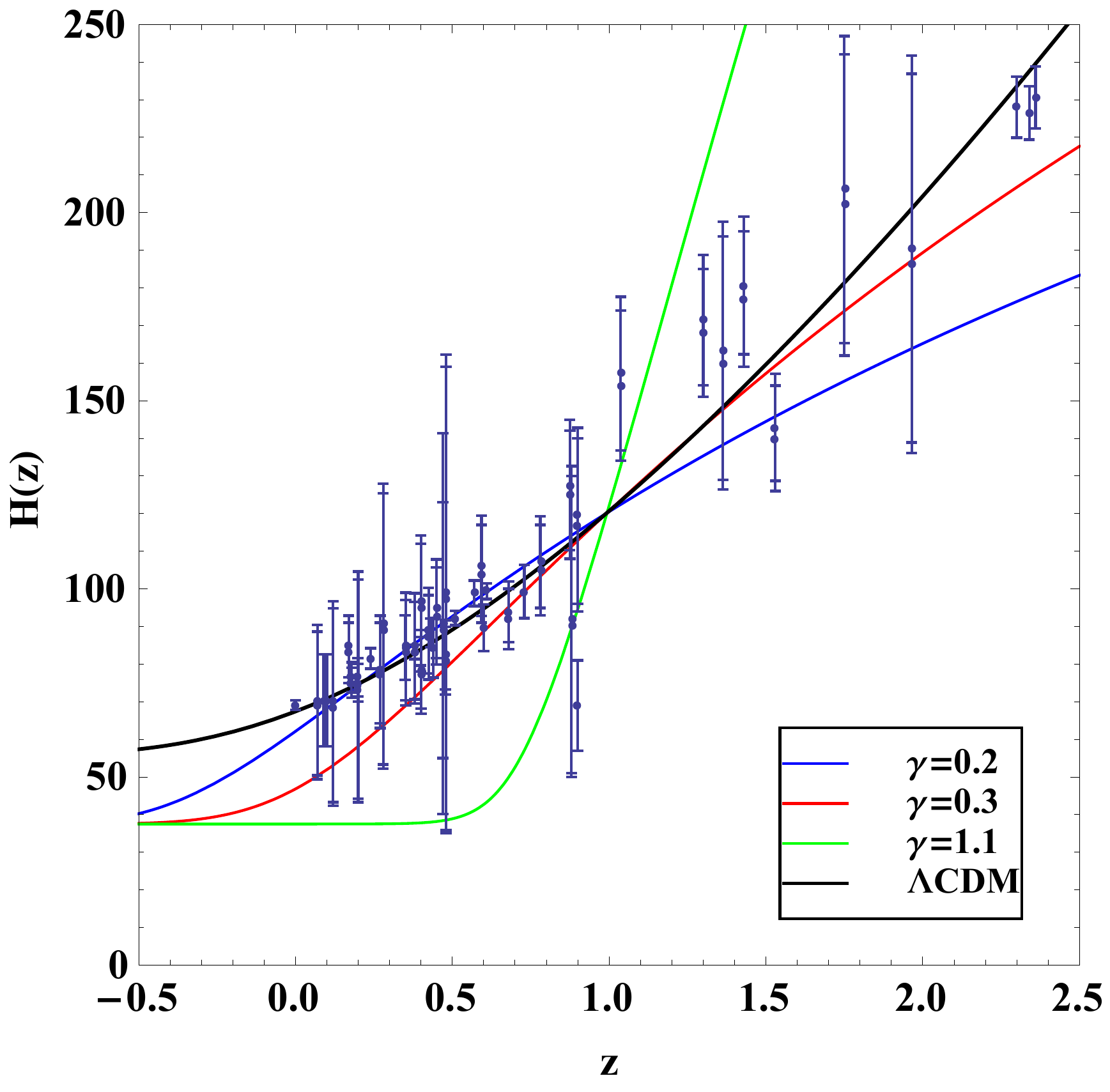}}\hfill
	\subfloat[]{\label{b}\includegraphics[scale=0.45]{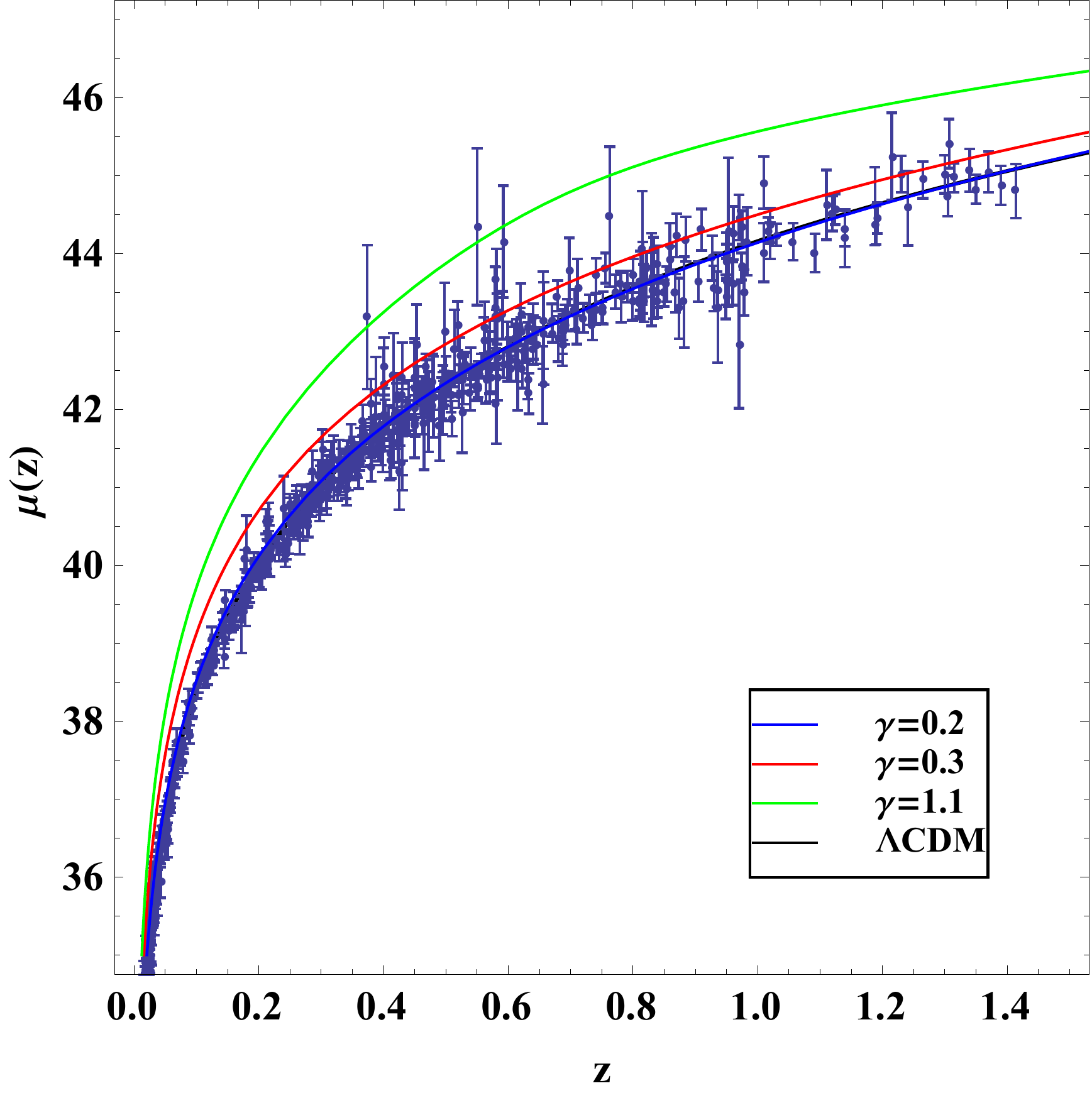}} 
	\caption{\scriptsize Error bar plots comparing our model with standard $ \Lambda $CDM using $ H(z) $ and $ SNeIa $ datasets respectively, with $ \alpha=0.05 $, $ \beta=-1 $. Our model is best fitted with $ \Lambda $CDM for $ \mu(z) $ data at present for all cases of $ \gamma $.} \label{Fig: Errorbarplots-z}
\end{figure}

The Hubble parameter $ H $ in terms of redshift $ z $ for the model is as follows:
	\begin{equation}\label{21}
		H(z) = \alpha \left(1+ProductLog\left[-\frac{\beta\left(\frac{a_0}{1+z}\right)^{-\frac{\gamma}{\alpha}}}{\alpha}\right]\right)
	\end{equation}
The distance modulus $ \mu(z) $ plotted in Fig. \ref{Fig: Errorbarplots-z}(b) is defined as
	\begin{eqnarray}\label{22}
		\mu(z) = m-M = 5 ~log_{10} D_L(z) + \mu_0,
	\end{eqnarray}
where $ m $ and $ M $ represent the apparent and absolute magnitudes of the object respectively and
\begin{equation}\label{23}
	D_L(z) = (1+z)c\int_0^z \frac{dz^*}{H(z^*)},  \mbox{\hspace{2em}}  \mu_0 = 25+5~log_{10}\left(\frac{c}{H_0}\right),
\end{equation}
are the luminosity distance and nuisance parameter respectively. Using the Planck 2018 results \cite{Planck:2018vyg}, we have $ H_0 = 67.4 ~km/s/Mpc $. 
In Fig. \ref{Fig: Errorbarplots-z}(a), the present model has been fitted using 77 points of $ H(z) $ dataset \cite{Shaily:2022enj} and union 2.1 compilation observational dataset \cite{SupernovaCosmologyProject:2011ycw} of 580 points has been used to fit the present model in Fig. \ref{Fig: Errorbarplots-z}(b). The Fig. \ref{Fig: Errorbarplots-z}(a) is not well fitted so much but the Fig. \ref{Fig: Errorbarplots-z}(b) is better fitted when we compare our model with $ \Lambda $CDM using $ H(z) $ and $ SNeIa $ data respectively. 

\subsection{ Statefinder diagnostic}
The statefinder diagnostic is a geometrical analysis to describe the phenomena of distinct dark energy (DE) models by the parameters $ r $ and $ s $, which are defined as
\begin{eqnarray}\label{24}
r=\frac{\dddot{a}}{aH^{3}},		\mbox{\hspace{2em}}		s=\frac{r-1}{3(q-\frac{1}{2})},
\end{eqnarray}
where $ q\neq \frac{1}{2} $.

The sequence of parameters $ \lbrace s,r \rbrace$ was proposed by Sahni et al. \cite{Sahni:2002fz,Sahni:2002yq} and Alam et al. \cite{Alam:2003sc}, where the variable represented by $ r $ is identical to the jerk parameter $ j $. A specific linear combination of the jerk parameter and the deceleration parameter $ q $ is used as another variable known as statefinder and represented by $ s $. For the value of scale factor $ a $ in Eq. (\ref{20}), the parameters $ r $ and $ s $ can be expressed in terms of $ t $ as

\begin{eqnarray}\label{25}
r=\frac{\alpha ^3 e^{3\gamma t}-\beta^3-3\beta^2\left(\gamma-\alpha\right)e^{\gamma t}-\beta\left(3\alpha^2-3\alpha\gamma+\gamma^2\right)e^{2\gamma t}}{\left(\alpha e^{\gamma t}-\beta\right)^3},	\nonumber	\\ 
s=\frac{2\beta\gamma e^{\gamma t} \left(3\beta+\left(\gamma-3\alpha\right)e^{\gamma t}\right)}{3\left(\alpha e^{\gamma t}-\beta\right) \left(3\alpha^2 e^{2\gamma t}+3\beta^2+2\beta\left(\gamma-3\alpha\right)e^{\gamma t}\right)}.
\end{eqnarray}

\begin{figure}\centering
	\subfloat[]{\label{a}\includegraphics[scale=0.4]{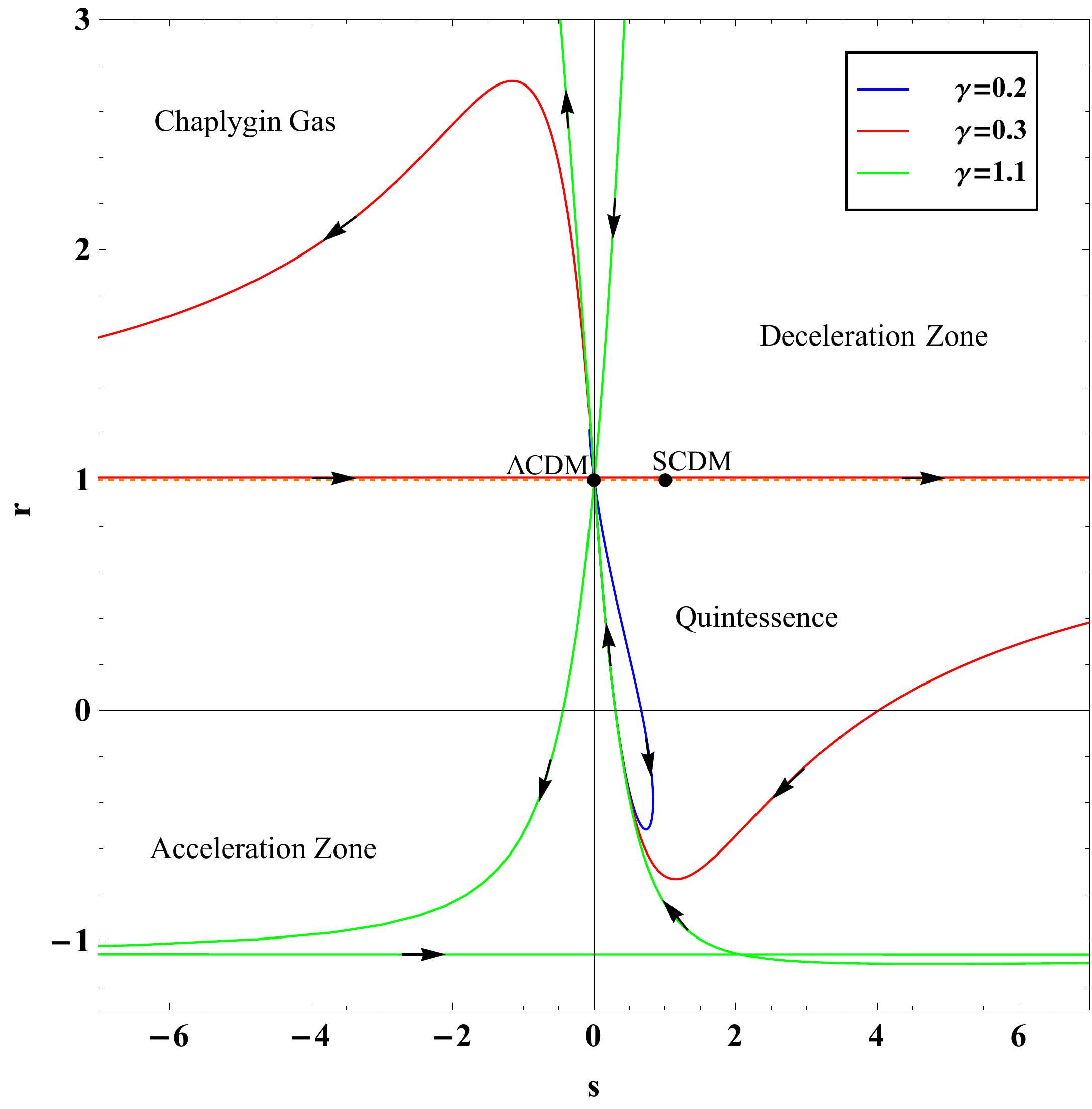}}\hfill
	\subfloat[]{\label{b}\includegraphics[scale=0.405]{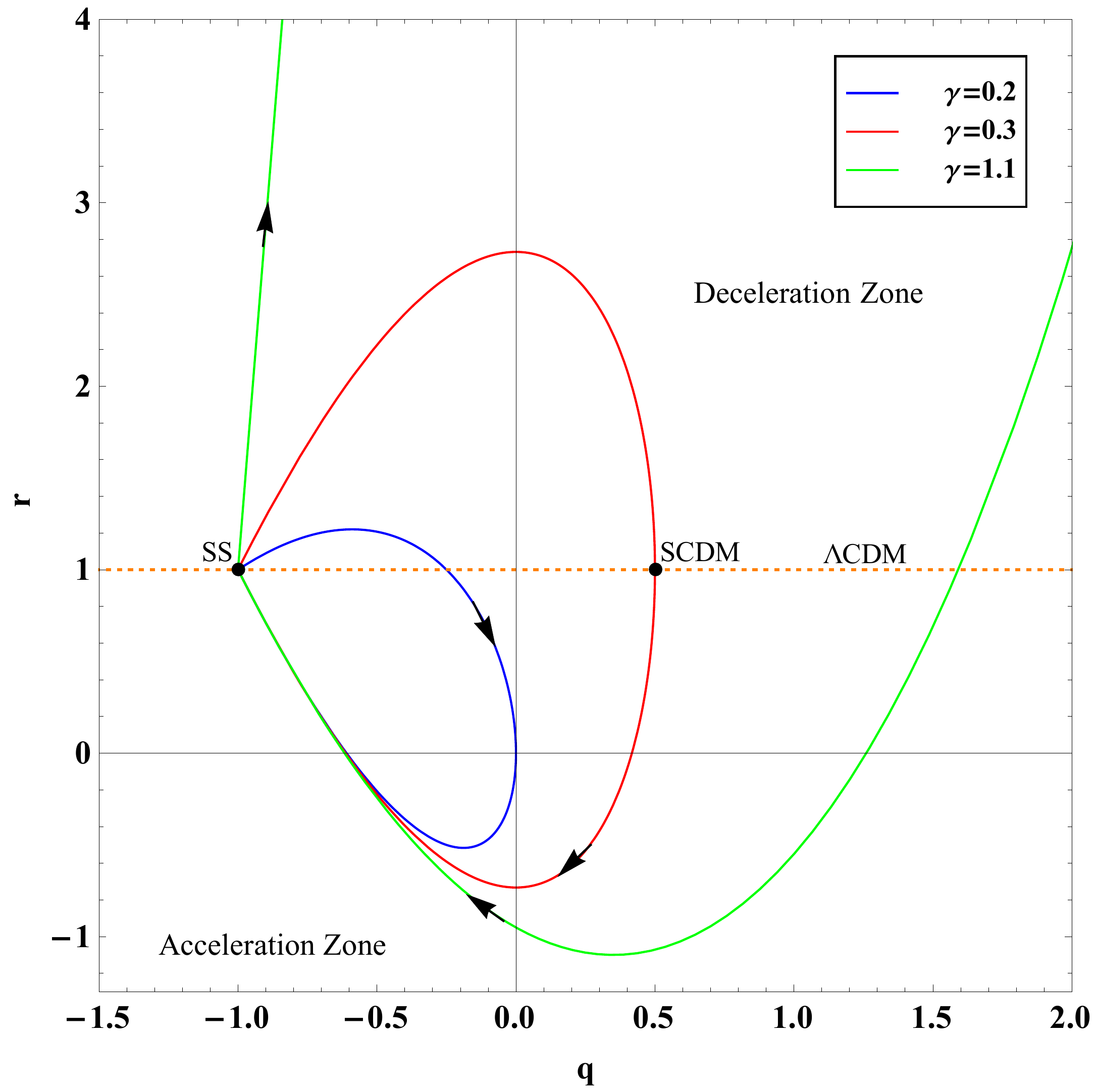}} 
	\caption{\scriptsize Depiction of $ s-r $ and $ q-r $ trajectories with $ \alpha=0.05 $ and $ \beta=-1 $.} \label{Fig:Statefinder}
\end{figure}

The $ r $ and $ s $ parameters are plotted in the three trajectories in the $ s-r $ plane for all values of $ \gamma $ as shown in Fig. \ref{Fig:Statefinder}(a). The arrows represent the directions of the $ s-r $ trajectories in the figure. The trajectory for $ \gamma=0.2 $ starts from the Chaplygin gas region ($ r>1,s<0 $), proceeds through the quintessence region ($ r<1, s>0 $) and then approaches the $ \Lambda $CDM ($ r=1,s=0 $). The trajectory for $ \gamma=0.3 $ starts from the Chaplygin gas region, overlaps the line $ r=1 $ and progresses through the quintessence region, and converges to $ \Lambda $CDM finally. The curve for $ \gamma=1.1 $ begins from the region $ r>1, s>0 $, moves through the quintessence region, crosses the $ \Lambda $CDM point and passes through the Chaplygin gas region before tracing the same path again. 

In the $ q-r $ plane, Fig. \ref{Fig:Statefinder}(b) depicts the three trajectories with time for all cases of $ \gamma $. The trajectory for $ \gamma=0.2 $ completes its journey in the region of $ q<0 $. The trajectories for $ \gamma=0.3 $ and $ 1.1 $ begin in $ q<0 $, move across the region $ q>0 $, and eventually complete their evolution in the $ q<0 $ region. The $ \gamma=0.3 $ trajectory also passes through the SCDM ($ r=1,q=1/2 $). These trajectories cross the $ \Lambda $CDM line, which is the line $ r=1 $ parallel to the $ q $-axis in the $ q-r $ plane. All the trajectories converge to steady state model (SS) ($ r=1,q=-1 $) at the end. Thus, our model behaves like the steady state model  based on the evolution of the trajectories to SS. 

\subsection{ Scalar Field Correspondence}
\begin{figure}\centering
	\subfloat[]{\label{a}\includegraphics[scale=0.55]{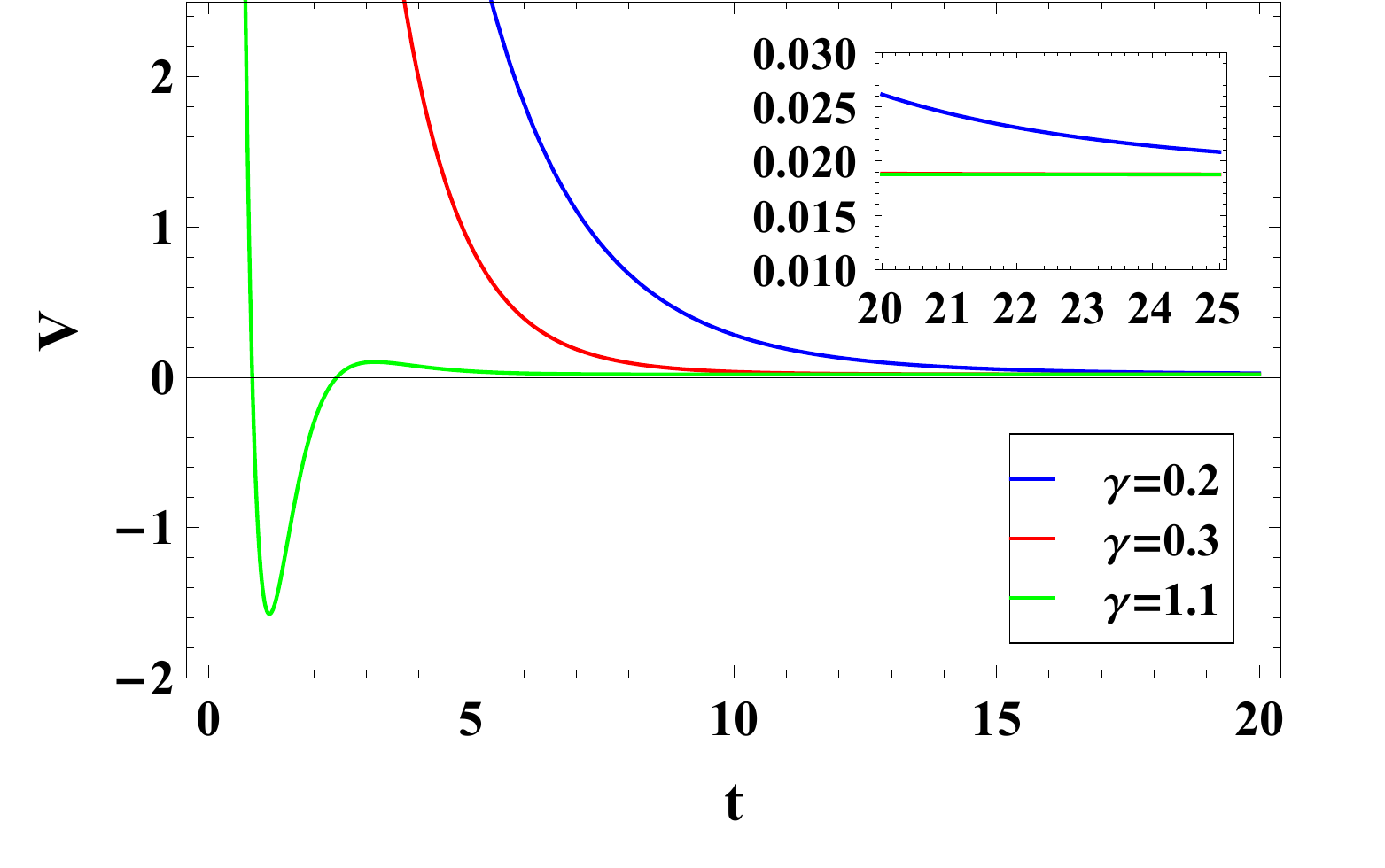}}
	\caption{\scriptsize Depiction of graph of scalar potential vs. $ t $ with $ \alpha=0.05 $, $ \beta=-1 $, $ m=0.6 $ and $ n=0.7 $.} \label{Fig: Potential-t}
\end{figure}

The scalar potential $ V\left(\phi\right) $ does not depend on $ \epsilon $ as given in Eq. (\ref{16}). Now, it is very difficult to express the scalar field $ \phi $ in closed form, so the potential function $ V\left(\phi\right) $ can not be expressed in terms of $ \phi $ explicitly. In Fig. \ref{Fig: Potential-t} the graph of scalar potential with time for some particular values of parameters shows that $ V\left(\phi\right) $ remains positive and decreasing for $ \gamma=0.2 $, $ 0.3 $, and settles to a positive constant in late times of the evolution of the universe. For $ \gamma=1.1 $, the scalar potential $ V\left(\phi\right) $ approaches a positive constant as $t\rightarrow\infty$, but it deviates to the region of negative potential for a short span of time at the initial stage of the evolution. 
\subsubsection{ Quintessence Scalar Field $\left(\epsilon=+1\right)$}
\begin{figure}\centering
	\subfloat[]{\label{a}\includegraphics[scale=0.5]{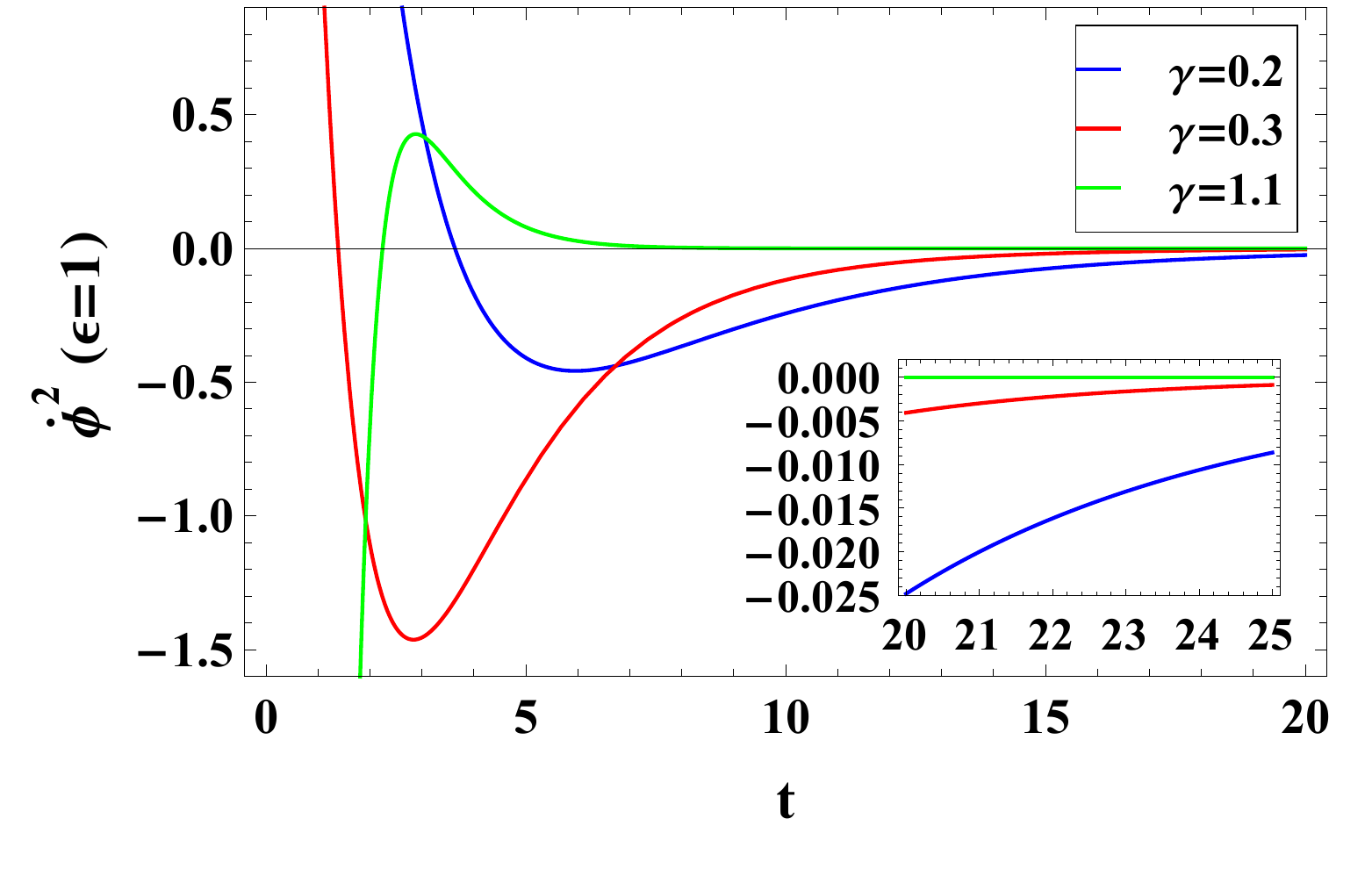}}\hfill
	\subfloat[]{\label{b}\includegraphics[scale=0.5]{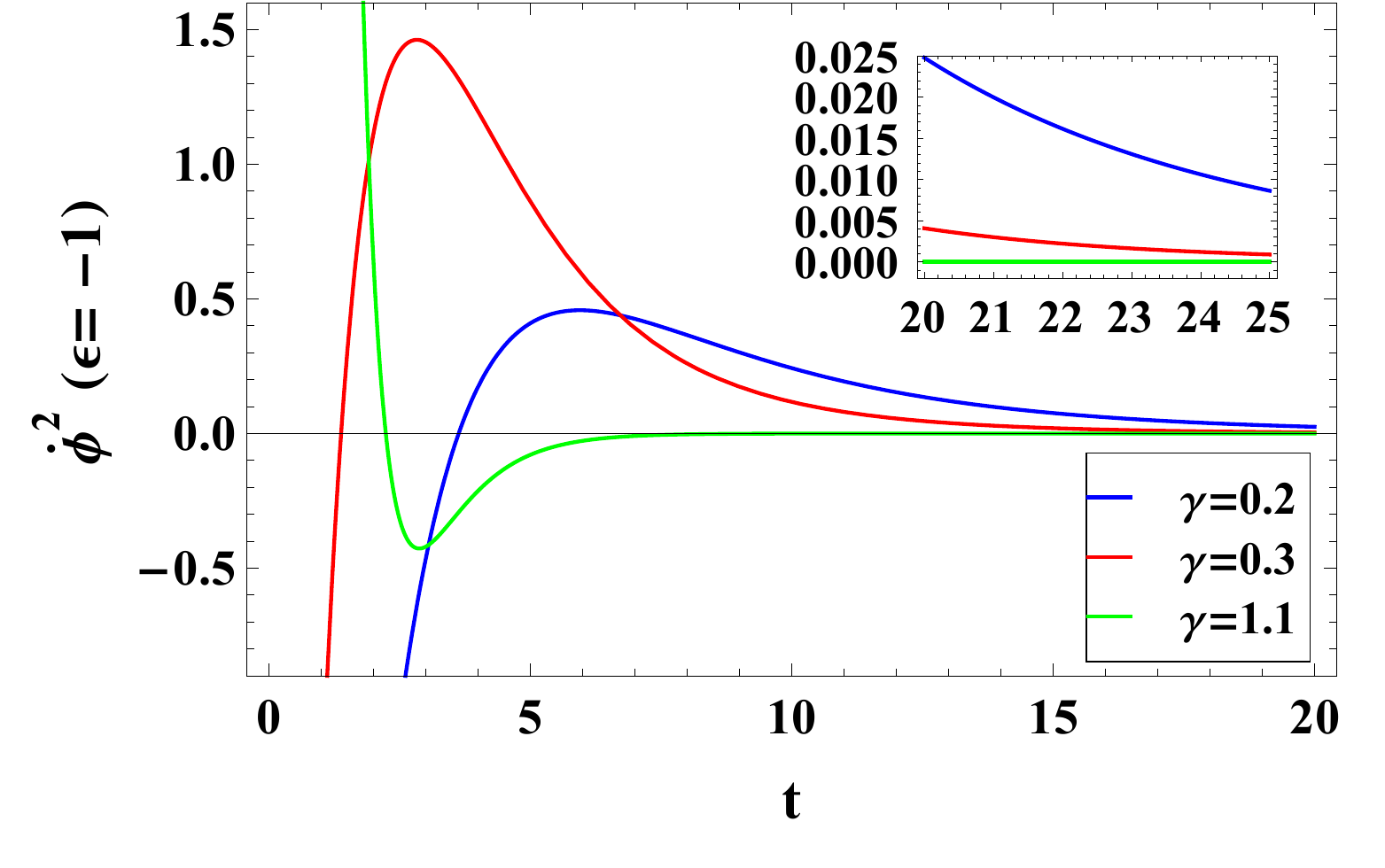}} 
	\caption{\scriptsize Depiction of graphs of $ \dot{\phi}^2 $ for $ \epsilon = 1 $ and $ -1 $ respectively vs. $ t $ with $ \alpha=0.05 $, $ \beta=-1 $, $ m=0.6 $ and $ n=0.7 $.} \label{Fig: Kinetic-t}
\end{figure}

In the quintessence scalar field, Eq. (\ref{16}) yields
\begin{equation}\label{26}
\dot{\phi}^2 = \frac{-2\dot{H}+12m\left(4\dot{H}H^2+3H\ddot{H}+6\dot{H}^2+\dddot{H}\right)}{n-1}.
\end{equation}
We observe that $ \dot{\phi}^2 $ is negative during the evolution in the late universe for $ \gamma=0.2, 0.3 $ as shown in Fig. \ref{Fig: Kinetic-t}(a) and hence $ \phi $ becomes imaginary. Thus, the kinetic term $ \left(\frac{1}{2}\dot{\phi}^2\right) $ is negative for the quintessence scalar field in the late times, but it must be positive. Therefore, the massless scalar field $ \phi $ can not be evaluated in cases of $ \gamma=0.2, 0.3 $. However, $ \dot{\phi}^2 $ is negative only for a short interval of time at the initial stage of the evolution for $ \gamma=1.1 $, and then its value increases suddenly and eventually decreases to zero in late times.
\subsubsection{ Phantom Scalar Field $\left(\epsilon=-1\right)$}
In the phantom scalar field, Eq. (\ref{16}) yields
\begin{equation}\label{27}
\dot{\phi}^2 = \frac{-2\dot{H}+12m\left(4\dot{H}H^2+3H\ddot{H}+6\dot{H}^2+\dddot{H}\right)}{-(n-1)}.
\end{equation}
We observe that the kinetic energy $ \dot{\phi}^2 $ is valid for $ \gamma=0.2 $, $ 0.3 $ as shown in Fig. \ref{Fig: Kinetic-t}(b) except for a very short time interval at the initial stage of the evolution. The value of $ \dot{\phi}^2 $ increases in the beginning but decreases to zero in late times for $ \gamma=0.2, 0.3 $. In contrast, except for a brief period, $ \dot{\phi}^2 $ is negative during the evolution initially and eventually converges to zero for $ \gamma=1.1 $.

\section{ Conclusions}\label{Conclusions section} 

\qquad In this paper, we consider the isotropic homogeneous cosmological model, which contains a massless scalar field $ \phi $ obeying a potential $ V(\phi) $. We have taken specific values for the arbitrary constants $ \alpha $, $ \beta $, $ \gamma $, $ m $ and $ n $. The graphical representations of various parameters are drawn after taking $ \alpha=0.05 $, $ \beta=-1 $, $ \gamma=0.2, 0.3 $ and $ 1.1 $, $ m=0.6 $, $ n=0.7 $. We summarise our work for this model as follows:
	\begin{itemize}
\item The model has no space-time singularity as the expansion scale factor $ a $ evolves from a finite size in the infinite past and increases exponentially with cosmic time $ t $.
	
\item The parametrized Hubble parameter $ H $ shows expansion forever and the future scenario of the universe attains Big Freeze.

\item The deceleration parameter $ q $ exhibits inflation at the initial stage of evolution for all values of $ \gamma $ and transition from the early decelerated expanding phase to the late time accelerated expanding phase for  $ \gamma =0.3,\,1.1 $. Also, the model shows eternal acceleration in the case of $ \gamma=0.2 $.
		
\item The model starts with a finite positive value of the energy density of massless scalar field $ \rho_\phi $ and eventually decreases monotonically to a finite positive quantity in late times, except for some short intervals of time at the initial stage. Also, the model starts with a finite negative value of the pressure of massless scalar field $ p_\phi $ and eventually monotonically increases to a finite negative quantity in late times, except for some short intervals of time at the initial stage.
		
\item From Fig. \ref{Fig:rhopomega-t}(c), we observe that the model shows highly unstable behaviour at the initial epoch for some values of $ \gamma $, but the model shows consistency with $ \Lambda $CDM in the late times for all values of $ \gamma $.

\item Our model is well fitted at the redshift $ z\simeq1 $ only, but it deviates for the values other than $ z\simeq1 $ when we compare it with the $ \Lambda $CDM using the $ H(z) $ dataset in the error bar plots. The model is well fitted with the $ \Lambda $CDM using the $ SNeIa $ dataset for $ \gamma = 0.2 $, but it deviates from the $ \Lambda $CDM for the values of redshift $ z $ that are not close to $ 0 $ in the cases of $ \gamma=0.3 $ and $ 1.1 $.

\item The plotted statefinder diagnostic shows that the $ s-r $ trajectories for $ \gamma < 1 $ start from the Chaplygin gas region, and all three trajectories pass through the $\Lambda $CDM model. All three trajectories in the $ q-r $ plane in the form of loops pass through the point, indicating the steady state (SS) model. In both the graphs, the trajectory for $ \gamma=0.3 $ passes through the SCDM, while the other trajectories deviate from it.
		
\item The massless scalar potential $ V(\phi) $ begins with a high potential at the initial stage of evolution and gradually decreases until it approaches a very small finite value for $ \gamma < 1 $, but for $ \gamma > 1 $, the potential $ V(\phi) $ begins with a high potential and gradually decreases until it approaches its minimum value, then increases up to a certain positive value, and finally approaches a finite value after the gradual decrement. For the case $ \gamma =1.1 $, the massless scalar potential $ V(\phi) $ becomes negative for a short period of time. 		
		
\item The quintessence-like scalar field kinetic term $ \dot{\phi}^2 $ violates all over the time except for some intervals at the beginning of the cosmic evolution for $ \gamma < 1 $, but in the case of $ \gamma > 1 $, it is valid throughout the evolution except for some intervals at the beginning. On the other hand, for the same values of $ \gamma $, the phantom-like scalar field kinetic term $ \dot{\phi}^2 $ behaves in the exact opposite manner of the quintessence-like scalar field kinetic term $ \dot{\phi}^2 $.

\end{itemize}

Therefore, in this work, we conclude that our model in $ f(R,T^\phi) $ gravity is a well explained model with the massless self-interacting scalar field $ \phi $. The model has no space-time singularity. The model is accelerated expanding and attains the Big Freeze scenario in the future evolution of the universe. The model predicts the moderate inflationary scenario at the time of the evolution of the universe, and it is consistent with $ \Lambda $CDM in late times. Our model turns out viable at present-day and in late times observations of $ H(z) $ and $ SNeIa $ dataset. Hence, we can say that our model plays an important role to describe the early as well as late times of the evolution of the universe.

\end{document}